\documentstyle[prl,aps,epsfig,twocolumn]{revtex}
\newcommand{\be}{\begin{equation}}
\newcommand{\ee}{\end{equation}}
\newcommand{\bea}{\begin{eqnarray}}
\newcommand{\eea}{\end{eqnarray}}

\newcommand{\p}{\partial}

\def\nn{\nonumber\\}

\begin{document}
\draft
\title{Spectral function of a quarter-filled one-dimensional
CDW insulator}

\author{Fabian H.L. Essler$^1$ and  Alexei M. Tsvelik$^{2}$}
\address{$^1$ Department of  Physics, University of Warwick,
Coventry CV4 7AL, UK\\
$^2$ Department of  Physics, Brookhaven National Laboratory, Upton, NY
11973-5000, USA}
\maketitle

\begin{abstract}
\par
We consider a one-dimensional charge density wave (CDW) insulator
formed by Umklapp processes in a quarter-filled band. The spectrum of
the model consists of gapless, uncharged excitations carrying spin
$\pm 1/2$ (spinons) and gapped, spinless excitations carrying charge
$\mp e/2$ (solitons and antisolitons).
We calculate the low-energy behaviour of the single-electron Green's
function at zero temperature. The spectral function exhibits a
featureless scattering continuum of two solitons and many spinons. 
The theory predicts that the gap observed by Angle Resolved
Photoemission (ARPES) is twice the activation gap in the dc
conductivity. We comment on possible applications to 
${\rm PrBa_2Cu_3O_7}$ and to the Bechgaard salts.

\end{abstract}


\narrowtext
\sloppy

The TMTSF and TMTTF families of quasi one dimensional (1D) organic
conductors have attracted much attention over the last 30 years. They
feature a rich variety of 3D ordered ground states, including spin and
charge density waves, spin-Peierls insulators and superconductors. 
Recently much of the attention has focussed on the 1D phases 
\cite{review,giam,vescoli} at low temperatures, which are believed to 
realize exotic strongly correlated Mott or CDW insulating states.
With regard to a  theoretical description of these phases there are
two main points of controversy. Firstly, it is not clear how
``one-dimensional'' in particular the TMTSF salts are in the
relevant regime of energies/temperatures, and secondly it is unknown
how strong the effective electron-electron interaction at low energies
is. Persuasive arguments in favour of strong interactions, giving rise
to a small value of the Luttinger liquid parameter $K_c$, are given in
\cite{review,giam,vescoli}. The authors of \cite{giam} quote an
estimate of $K_c \approx 0.22$. At such values of $K_c$ the 
single-particle tunneling between the chains should be strongly
suppressed, leading to the enhancement of 1D effects. 
Perhaps, the strongest evidence in favour of such a scenario comes
from ARPES measurements \cite{zwick,private}. Experiments done by
different groups not only show the absence of quasi-particle peaks,
but find no evidence of any dispersing features in the single-electron
spectral function. Quite similar behaviour has recently been observed
in the quarter-filled ${\rm Cu-O}$ chain material ${\rm PrBa_2Cu_3O_7}$
(``P123'') \cite{p123arpes,p123optics}.

The great difficulty in determining dynamical response functions
theoretically is the presence of Umklapp scattering processes, which
dynamically generate a spectral gap. In this work we use a
method based on the integrability of the low-energy theory to
calculate the single-particle Green's function for the first time. 
In the Bechgaard salts there are two separate mechanisms that lead to
Umklapp scattering: (1) ``double'' Umklapp processes due to the
commensurate band filling 1/4 \cite{4umklapp,yoshioka}. These generate
a gap only for strong interactions ($K_c<0.25$). 
(2) a small dimerization, which halves the Brillouin zone and gives
rise to ``single'' Umklapp processes \cite{vic,penc2}.
These open a gap already for weak interactions $K_c<1$ but their
coupling constant is proportional to the dimerization and thus small.
It is an open question which of these two mechanisms dominates in the 
Bechgaard salts. 
Here we develop a theory for ARPES in the case where one of the
processes can be neglected. We consider the case with only double
Umklapp scattering in detail and present some results for the
dimerized case at the end. Following \cite{4umklapp,review,giam} we
adopt a field theory description of the quarter-filled CDW insulator
at low energies. The small parameter is this approach is the ratio of
the gap to the bandwidth $W$. The underlying lattice model may be
thought of as a quarter-filled Hubbard model with additional
density-density interactions \cite{penc,nakamura,yoshioka}. The
Lagrangian density is the sum of two terms describing the spin and
charge degrees of freedom respectively (``spin-charge separation'')
${\cal L} = {\cal L}_s + {\cal L}_c$, where
\bea
{\cal L}_s &=& \frac{1}{16\pi}\left[v_s^{-1}(\p_{\tau}\Phi_s)^2 +
v_s(\p_x\Phi_s)^2\right], \nn 
{\cal L}_c &=& \frac{1}{16\pi}[v_c^{-1}(\p_{\tau}\Phi_c)^2 +
v_c(\p_x\Phi_s)^2] +  \lambda\cos(\beta\Phi_c)\ .
\label{lagr}
\eea
Here $\beta=\sqrt{4K_c}$ and the spin sector is gapless
\cite{backscatt} and spin rotationally invariant, which fixes the
corresponding Luttinger liquid parameter. 
The cosine term in the charge sector is generated by double Umklapp
processes and we assume it to be relevant, which implies $\beta < 1$. 
We will take $\beta^2 \approx 0.9$ which best fits the estimates given
in \cite{giam} for the case of Bechgaard salts. The spin and charge
velocities $v_s$ and $v_c$ are parameters of the theory.

\begin{figure}[ht]
\begin{center}
\epsfxsize=0.35\textwidth
\epsfbox{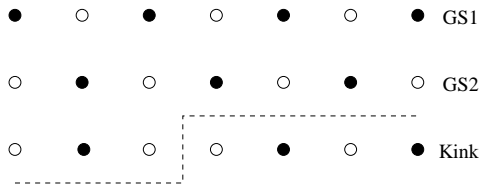}
\end{center}
\caption{\label{fig:kink}
Ground states and antisolitonic kink in a $U$-$V$ extended Hubbard
model. 
}
\end{figure}

The electronic creation and annihilation operators
$c^\dagger_{j,\sigma},c_{j,\sigma}$ ($\sigma=\uparrow,\downarrow$)
of the underlying lattice model are related to the fields appearing in
the Lagrangian and their corresponding {\sl dual fields}
\be
\Theta_{c,s}(x)=\frac{-i}{v_{c,s}}\int_{-\infty}^xdy\
\partial_\tau\Phi_{c,s}(\tau,y)
\ee
by
\begin{equation}
c_{j,\sigma}\propto\left[\exp(ik_Fx)\ R_\sigma(x)+
\exp(-ik_Fx)\ L_\sigma(x)\right]\ ,
\end{equation}
where $x=ja_0$ ($a_0$ is the lattice spacing) and
\bea
L_{\sigma} &=&\eta_{\sigma}\ 
e^{\frac{i}{4}\left(\frac{\beta}{2}\Phi_c-\frac{2}{\beta}\Theta_c\right)}\ 
e^{\frac{i}{4}\tilde{\sigma}(\Phi_s-\Theta_s)},\nn
R_{\sigma} &=&\eta_{\sigma}\
e^{-\frac{i}{4}\left(\frac{\beta}{2}\Phi_c+\frac{2}{\beta}\Theta_c\right)}\ 
e^{-\frac{i}{4}\tilde{\sigma}(\Phi_s+\Theta_s)}.
\label{RL}
\eea
Here $\tilde{\sigma}=\pm$ for up/down spins and
$\eta_{\sigma}=\eta_\sigma^\dagger$ are Klein factors that fulfil 
$\lbrace\eta_\sigma,\eta_\tau\rbrace=2\delta_{\sigma,\tau}$.
In order to determine the spectral function at zero temperature we
evaluate the single-electron Green's function
\be
G^{(\sigma)}_{RR}(\tau,x)=\langle
0|R_\sigma(\tau,x)R^\dagger_\sigma(0)|0\rangle . 
\ee
We do this by exploiting the factorization (\ref{RL})
\bea
G^{(\sigma)}_{RR}(\tau,x)&=&{_c\langle} 0|{\cal O}^\dagger_c(\tau,x)\
{\cal O}_c(0)|0\rangle_c\nn
&\times& {_s\langle} 0|
e^{\mp\frac{i}{4}(\Phi_s+\Theta_s)(\tau,x)}\
e^{\pm\frac{i}{4}(\Phi_s+\Theta_s)(0)}
|0\rangle_s\ ,
\label{fact}
\eea
where we have introduced the shorthand notation ${\cal O}_{c}$ for
the charge part of $R_\sigma$ in (\ref{RL}). The subscripts $c,s$
label the vacua in the charge and spin sector respectively.
The spin part of (\ref{fact}) is easily evaluated 
as the spin sector in (\ref{lagr}) is a free bosonic theory
\bea
G^{(\sigma)}_{RR}(\tau,x)&=&{_c\langle} 0|{\cal O}^\dagger_c(\tau,x)\
{\cal O}_c(0)|0\rangle_c\ (v_s\tau-ix)^{-\frac{1}{2}}.
\eea
The large-distance asymptotics of the correlation function in the
charge sector can be analyzed by going to the spectral representation
of the corresponding sine-Gordon model (SGM). In the repulsive regime
$\beta>1/\sqrt{2}$, the only single-particle excitations of the
SGM (\ref{lagr}) are soliton and antisoliton. A rough physical picture
of what solitonic excitations look like may be obtained by considering a
quarter filled extended Hubbard model with sufficiently large
nearest-neighbour repulsion $V$ \cite{penc,yoshioka}, so
that we are in the $4k_F$ CDW insulating phase. One
naively obtains two ground states as shown in Fig.\ref{fig:kink} and
the antisoliton is then the kink connecting these two ground states.
It is also clear from Fig.\ref{fig:kink}, the soliton carries an
effective charge of $-{e}/{2}$. Soliton and antisoliton have massive
relativistic dispersions, which we as usual parametrize in terms of a
rapidity variable $\theta$ 
\be
E(\theta)=M\cosh\theta\ ,\quad P(\theta)=\frac{M}{v_c}\sinh\theta
=m\sinh\theta\ .
\ee
A basis of scattering states of 
solitons and antisolitons can be constructed by means of the
Zamolodchikov-Faddeev (ZF) algebra. The ZF algebra is essentially
the extension of the algebra of creation and annihilation operators
for free fermions or bosons to the case or interacting particles with
factorizable scattering. The ZF algebra is based on the knowledge of
the exact spectrum and scattering matrix \cite{SM}. For
the SGM the ZF operators (and their hermitian
conjugates) satisfy the following algebra
\begin{mathletters}
\label{fz1}
\begin{eqnarray}
{Z}^{\epsilon_1}(\theta_1){Z}^{\epsilon_2}(\theta_2) &=&
S^{\epsilon_1,\epsilon_2}_{\epsilon_1',\epsilon_2'}(\theta_1 -
\theta_2){Z}^{\epsilon_2'}(\theta_2){Z}^{\epsilon_1'}(\theta_1)\; ,
\\
{Z}_{\epsilon_1}^\dagger(\theta_1)Z_{\epsilon_2}^\dagger(\theta_2) &=&
Z_{\epsilon_2'}^\dagger(\theta_2){ Z}_{\epsilon_1'}^\dagger
(\theta_1)S_{\epsilon_1,\epsilon_2}^{\epsilon_1',\epsilon_2'}(\theta_1 -
\theta_2)\; ,\\
Z^{\epsilon_1}(\theta_1)Z_{\epsilon_2}^\dagger(\theta_2) 
&=&Z_{\epsilon_2'}
^\dagger(\theta_2)
S_{\epsilon_2,\epsilon_1'}^{\epsilon_2',\epsilon_1}
(\theta_2-\theta_1)Z^{\epsilon_1'}(\theta_1)\nonumber\\
&&+(2 \pi) \delta_{\epsilon_2}^{\epsilon_1} 
\delta (\theta_1-\theta_2),
\end{eqnarray}\end{mathletters}%
where $S^{\epsilon_1,\epsilon_2}_{\epsilon_1',\epsilon_2'}(\theta)$ are
the known factorizable two-particle scattering matrices \cite{SM} and
$\varepsilon_j=-,+$ for solitons and antisolitons respectively.
Using the ZF operators a Fock space of states can be constructed as
follows. The vacuum is defined by
\begin{equation}
Z_{\varepsilon_i}(\theta) |\Omega\rangle=0 \ .
\end{equation}
Multiparticle states are then obtained by acting with strings of
creation operators $Z_\epsilon^\dagger(\theta)$ on the vacuum
\begin{equation}
|\theta_n\ldots\theta_1\rangle_{\epsilon_n\ldots\epsilon_1} = 
Z^\dagger_{\epsilon_n}(\theta_n)\ldots
Z^\dagger_{\epsilon_1}(\theta_1)|\Omega\rangle . 
\label{states}
\end{equation} 
In term of this basis the resolution of the identity is given by
\begin{eqnarray}
&& \openone = |\Omega\rangle\langle \Omega| \label{identity} \nn
&& + \sum_{n=1}^\infty\sum_{\epsilon_i}\int_{-\infty}^{\infty}
\frac{d\theta_1\ldots d\theta_n}{(2\pi)^nn!}
|\theta_n\ldots\theta_1\rangle_{\epsilon_n\ldots\epsilon_1}
{}^{\epsilon_1\ldots\epsilon_n}\langle\theta_1\ldots\theta_n|\ .
\nonumber
\end{eqnarray}
Using this basis of states in the charge sector SGM we obtain the
following spectral representation
\begin{eqnarray}
\label{corr}
&&{_c\langle} 0|{\cal O}^\dagger_c(\tau,x)\
{\cal O}_c(0)|0\rangle_c
=\sum_{n=1}^\infty\sum_{\epsilon_i}\int
\frac{d\theta_1\ldots d\theta_n}{(2\pi)^nn!}
\nonumber\\
&&\times
e^{-{\sum_{j} iP(\theta_j)x+E(\theta_j) \tau}}
|\langle \Omega| {\cal O}_c^\dagger(0,0)|\theta_n\ldots\theta_1
\rangle_{\epsilon_n\ldots\epsilon_1}|^2.
\end{eqnarray}

\begin{figure}[ht]
\begin{center}
\epsfxsize=0.4\textwidth
\epsfbox{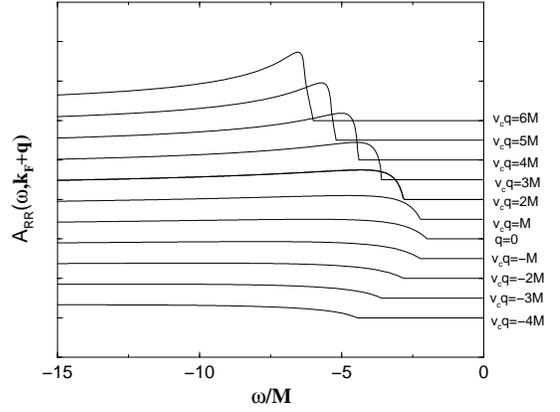}
\end{center}
\caption{\label{fig:constq}
$A_{RR}(\omega,k_F+q)$ as a function of $\omega/M$ for$v_s=0.8v_c$. The
curves for different $q$ are offset.
}
\end{figure}

The first non-vanishing matrix element (formfactor) in (\ref{corr})
corresponds to the emission of two solitons. According to Eq.(2.12) of
\cite{LukZam01} we have 
\bea
&&|{\langle \Omega}|{\cal O}^\dagger_c(0)|\theta_1,\theta_2{\rangle_{--}}|^2
= Z\ e^{-\frac{\theta_1 + \theta_2}{4}}|G(\theta_1-\theta_2)|^2,\nn
&&G(\theta)=iC_1 \sinh{\theta/2}\nn
&&\times \exp \left ( \int_0^\infty\frac{dt}{t}
\frac{\sinh^2(t[1-i\theta/\pi])\sinh(t[\xi-1])}{\sinh 2t\ \sinh\xi t\ \cosh t}\right),\nn
&&C_1=\exp\left(-\int_0^\infty \frac{dt}{t}
\frac{\sinh^2(t/2)\ \sinh(t[\xi-1])}{\sinh 2t\ \sinh\xi t\ \cosh t}\right),
\end{eqnarray}
The overall constant $Z$ is calculated in \cite{LukZam01} and is
fixed by the following {\sl short-distance} normalization for
$r=\sqrt{v_c^2\tau^2+x^2}\to 0$ 
\be
{_c\langle} 0|{\cal O}^\dagger_c(\tau,x)
{\cal O}_c(0)|0\rangle_c\longrightarrow\left(
\frac{v_c\tau+ix}{v_c\tau-ix}\right)^\frac{1}{4} 
r^{-\frac{\beta^2}{16}-\frac{1}{\beta^2}}.
\label{short}
\ee
We note that (\ref{short}) implies that $Z$ has dimensions
of $L^{\frac{\beta^2}{16}-\frac{1}{\beta^2}}$. In order to restore proper
units one needs to multiply the correlation function by an appropriate 
power of the lattice spacing. The first subleading contribution to the
correlation function involves three solitons and one antisoliton and
is expected to be small at low energies, so we do not consider it here. 
The contribution of intermediate states involving two solitons to the
single-electron Green's function is therefore given by 
\bea
&&G_{RR}^{(\sigma)}(\tau,x) \simeq \frac{Z}{\sqrt{(v_s\tau- i x)}}
\int_{-\infty}^{\infty}\frac{d\theta_-}{2\pi}|G(2\theta_-)|^2\nn
&&\times\int_{-\infty}^\infty \frac{d\theta_+}{2\pi} e^{-\theta_+/2}\ e^{-
2m\cosh\theta_-(v_c\tau\cosh\theta_+ + i x\sinh\theta_+)} \nn
&&=\frac{Z(2\pi)^{-1/2}}{\sqrt{(v_c\tau - i x)(v_s\tau - ix)}}
\int_{-\infty}^\infty \frac{d\theta}{{2\pi}}
\frac{e^{-2 mr\cosh\theta}|G(2\theta)|^2}{\sqrt{2m\cosh\theta}}.\nn
\eea
Fourier transforming and analytically continuing we find
\bea
&&G^{(\sigma)}_{RR}(\omega, k_F+q) =
-{Z}\sqrt{\frac{2v_c^2}{v_c+v_s}}\int_{-\infty}^\infty\frac{d\theta}{\sqrt{2\pi}}
\frac{|G(2\theta)|^2}{\sqrt{c(\theta)}}\nn
&&\times\frac{\omega+v_cq}{\sqrt{c^2(\theta)-s^2}}
\Biggl[\left(c(\theta)+\sqrt{c^2(\theta)-s^2}\right)^2
\!-\frac{(\omega+v_cq)^2}{\alpha}\Biggr]^{-\frac{1}{2}}\!\!,\nn
\label{G}
\eea
where $s^2=\omega^2-v_c^2q^2$, $c(\theta)=2M\cosh\theta$, $\alpha =
(v_c+v_s)/(v_c-v_s)$. In the limit $v_s=v_c=v$ (\ref{G}) simplifies to
\bea
&&G^{(\sigma)}_{RR}(\omega, k_F+q) =
\frac{Z\sqrt{v/2\pi}}{\omega -vq}\nn
&&\times\int_{-\infty}^{\infty}{d\theta}
\frac{|G(2\theta)|^2}{\sqrt{2M\cosh\theta}}
\left[1 - \frac{2M\cosh\theta}{\sqrt{(2M\cosh\theta)^2 -s^2}}\right].
\eea
The spectral function for $v_c=v_s$ is
\bea
&&A_{RR}(\omega,k_F+q) = \frac{\tilde{Z}}{\omega -
vq}
\int_{0}^b{d\theta}
\frac{|G(2\theta)|^2\ \sqrt{c(\theta)}
}{\sqrt{s^2- c^2(\theta)}}\ ,
\eea
where $b={\rm arccosh}(s/2M)$
and $\tilde{Z}={Z\sqrt{2v/\pi^3}}$. In Fig.\ref{fig:constq} we plot
$A_{RR}(\omega,k_F+q)$ as a function of $\omega$ for $v_s=0.8 v_c$ and
11 different values of $q$.
As is clear from Fig.\ref{fig:constq} the spectral function is rather
featureless and there are no singularities. Furthermore, in contrast
to the half-filled Mott insulator \cite{shen,ET}, there are no
dispersing features associated separately with with $v_c$ and $v_s$.
Just above the threshold at $s^2=\omega^2-v^2q^2=4M^2$ one
can approximate $|G(2\theta)|=|C_1\sinh\theta|$, so that for $v_c=v_s$
\bea
A_{RR}(\omega,q) \approx
Z\sqrt{\frac{v}{4\pi M}}\frac{1}{\omega -
vq}\left(\frac{s}{2M}-1\right). 
\eea

Thus the spectral weight increases linearly with $s-2M$ above the
threshold. There is a remarkable relation between the spectral
function and the real part of the optical conductivity
$\sigma(\omega)$, which for $\beta\to 1$ and $v_c=v_s$ takes the
simple form
\bea
A_{RR}(\omega,q) \approx 
\frac{{\rm const}}{\omega - q}\int_{2M}^{s}
dx\frac{x^{5/2}\sigma(x)}{\sqrt{\omega^2 - v^2q^2 - x^2}}. 
\eea
The tunneling density of states for $v_c=v_s$ is
\bea
\rho(\omega) = \frac{2Z}{\sqrt{vM}}\int_0^{{\rm arccosh}(\omega/2M)}
\frac{d\theta}{\sqrt{2\pi}} \frac{|G(2\theta)|^2}{\sqrt{2\cosh\theta}},
\eea
and displays a roughly linear increase after an initial
$(\omega-2M)^{3/2}$ behaviour just above the threshold at $\omega=2M$.

Let us now try to relate our results to experiments. The ARPES data for
P123 (Fig.3 of \cite{p123arpes}) shows a single, very broad,
dispersing feature that is asymmetric around $k_F$ and sharpens in the 
region $k>k_F$. If we interpret the underlying increase in intensity
in the data as background, the signal has a form quite similar to our
Fig.\ref{fig:constq}. It would be interesting to extract a value
$\Delta_{\rm PE}$ for the gap from the ARPES data and compare it to
gaps seen in optical measurements $\Delta_{\rm opt}$ and the thermal
activation gap $\Delta_{\rm T}$ extracted e.g. from the dc
conductivity. Our theory predicts $\Delta_{\rm PE}=\Delta_{\rm
opt}=2\Delta_{\rm T}$.

Our results may also be of relevance to the Bechgaard salts and in
particular to (TMTTF)$_2$PF$_6$, where 3D effects appear to be
small. However, the ratio of the observed gap to the bandwidth is not
small, so that one should proceed with caution when applying a
field-theoretical approach. Furthermore, we have neglected
dimerization, which generates an additional interaction $\mu
\cos\frac{\beta}{2}\Phi_c$ in the charge sector of (\ref{lagr}). 
The resulting model is a double SGM \cite{DSG} and not treatable by
the method used here. If the coupling $\lambda$ of the double Umklapp
is positive (as is the case for sufficiently strong nearest-neighbour
repulsion $V$ in the underlying extended Hubbard model) the
dimerization does not lead to a qualitative change of the physics 
described above. If $\lambda$ is negative, the dimerization leads to
the confinement of the kinks and we expect the physics to be more
similar to the case without double Umklapp described below. At present
no definite estimates for $U$ and $V$ for the Bechgaard salts are available.
Having said this, let us try to compare our results to the experimental
findings on (TMTTF)$_2$PF$_6$. The experimental estimates for the gaps are
$\Delta_{\rm PE} =100 \pm 20$ meV, $\Delta_{\rm opt} \approx 100$ meV
and $\Delta_{\rm T}\approx 43$ meV \cite{vescoli2,zwick}. This fits well
into the small-$K_c$ scenario discussed above: $\Delta_{\rm opt}$ is
{\sl twice} the single-particle gap because the lowest-lying
intermediate states that couple to the current operator are
soliton-antisoliton states \cite{controzzi}. In the quarter-filled
case, the gap observed in photoemission corresponds to the emission of
{\sl two} solitons and therefore also occurs at twice the
single-particle gap. This is a distinct feature of the quarter-filled
$4k_F$-CDW insulator and it looks like there is qualitative agreement
with the experiments here.
%

\begin{figure}[ht]
\begin{center}
\epsfxsize=0.4\textwidth
\epsfbox{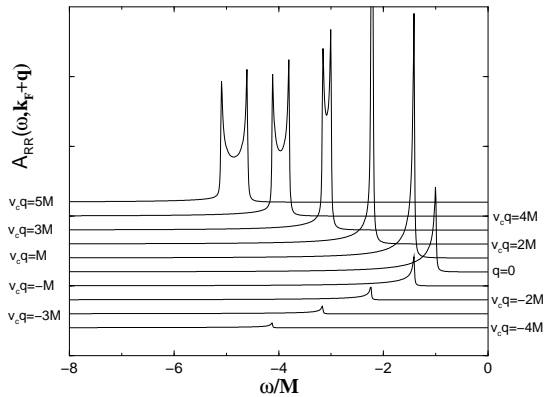}
\end{center}
\caption{\label{fig:a1}
$A_{RR}(\omega,k_F+q)$ as a function of $\omega/M$ with dimerization but
no double Umklapp and $v_s=0.8v_c$. The curves for different $q$ are 
offset and have been smoothed to remove the square root threshold
singularities. 
}
\end{figure}

Let us now turn to the case where we have dimerization, but the
interactions are weak so that the double Umklapp is irrelevant. The
relevant Lagrangian is again of the form (\ref{lagr}), but now
$\beta^2=K_c$. The main difference to the double-Umklapp case is the
expression for the electron creation operator, which is obtained by
replacing $\beta/2$ by $\beta$ in (\ref{RL}).
The spectral function can then be calculated by the same
methods as before \cite{ET}, but now solitons carry charge $e$ and
only a {\sl single} soliton couples to the electron creation operator
at low energies.  This means that now the gap $M$ seen in ARPES is the
same as the thermal activation gap and half the optical gap. 
In Fig.\ref{fig:a1} we show the spectral function for $v_s=0.8 v_c$
and 10 different values of $q$. 
For $v_c q\leq Q=Mv_s/\sqrt{v_c^2-v_s^2}$
The spectral function exhibits square root singularities above the
threshold at $\omega=\sqrt{M^2+v_c^2q^2}$.
For $v_c q> Q$ there are two square-root singularities: one above
the threshold at $\omega=v_sq+M\sqrt{1-(v_s/v_c)^2}$ and a second one
when $\omega$ approaches $\sqrt{M^2+v_c^2q^2}$ from below. Their
interpretation is the same as for the half-filled Mott insulator
\cite{shen}: the first singularity correponds to states where the
holon does not carry momentum, and the second one to states where the
spinon does not carry momentum.

This scenario appears to be much less compatible with the available
ARPES data than the double Umklapp scenario in the sense that it
predicts a rather sharp, dispersing feature with significant spectral
weight around the threshold. In the data essentially no dispersing
feature has been observed, which would appear to be more compatible
with the very broad signal shown in Fig. \ref{fig:constq} than with
the one shown in Fig. \ref{fig:a1}.

We are grateful to T. Valla,  P. D. Johnson and A. Chubukov for stimulating 
discussions.
This work was supported by the DOE under contract number DE-AC02-98 CH 10886 
(A.M.T) and by the EPSRC under grants AF/100201 and GR/N19359 (FHLE).


\begin{thebibliography}{99}
\bibitem{backscatt}
For some values of the parameters of the underlying
extended Hubbard model the spin sector can be gapped. This case can be
treated by the same methods.
\end{thebibliography}
\end{document}